\documentclass[10pt,conference,hidelinks]{IEEEtran}\IEEEoverridecommandlockouts
\usepackage{epsfig,graphicx,subfigure,psfrag,amsmath,cases}
\usepackage{latexsym,amssymb,amsmath,epsfig,subfigure,algorithm,mathtools}
\usepackage{algorithmic}
\usepackage{color}
\usepackage{url}
\usepackage{scrtime}
\usepackage{cite}
\usepackage{epstopdf}
\usepackage{subfigure}
\usepackage{bbding}
\usepackage{multicol}
\usepackage{bm}
\usepackage{xcolor}
\usepackage[framemethod=TikZ]{mdframed}
\usetikzlibrary{shadows}
\usepackage{environ}
\usepackage{varwidth}

\newlength{\MyMdframedWidthTweak}%
\NewEnviron{MyMdframed}[1][]{%
    \setlength{\MyMdframedWidthTweak}{\dimexpr%
        +\mdflength{innerleftmargin}
        +\mdflength{innerrightmargin}
        +\mdflength{leftmargin}
        +\mdflength{rightmargin}
        }%
    \savebox0{%
        \begin{varwidth}{\dimexpr\linewidth-\MyMdframedWidthTweak\relax}%
            \BODY
        \end{varwidth}%
    }%
    \begin{mdframed}[
        backgroundcolor=lightgray,
        shadow=true,
        shadowsize=4pt,
        roundcorner=5pt,
        userdefinedwidth=\dimexpr\wd0+\MyMdframedWidthTweak\relax,
        #1]
        \usebox0
    \end{mdframed}
}

\author{Zhiqiang Wei, Weijie Yuan, Shuangyang Li, Jinhong Yuan, and Derrick Wing Kwan Ng
	\thanks{Zhiqiang Wei, Weijie Yuan, Shuangyang Li, Jinhong Yuan, and Derrick Wing Kwan Ng are with the School of Electrical Engineering and Telecommunications, the University of New South Wales, Australia (email: zhiqiang.wei; weijie.yuan; shuangyang.li; j.yuan; w.k.ng@unsw.edu.au).} }

\title{Performance Analysis and Window Design for Channel Estimation of OTFS Modulation}

\newtheorem{T-Prob}{Transformed Problem}

\newcommand{\abs}[1]{\lvert#1\rvert}

\begin{document}
\maketitle
\begin{abstract}
In this paper, we investigate the impacts of transmitter and receiver windows on orthogonal time-frequency space (OTFS) modulation and propose a window design to improve the OTFS channel estimation performance.
Assuming ideal pulse shaping filters at the transceiver, we first identify the role of window in effective channel and the reduced channel sparsity with conventional rectangular window.
Then, we characterize the impacts of windowing on the effective channel estimation performance for OTFS modulation.
Based on the revealed insights, we propose to apply a Dolph-Chebyshev (DC) window at either the transmitter or the receiver to effectively enhance the sparsity of the effective channel.
As such, the channel spread due to the fractional Doppler is significantly reduced, which leads to a lower error floor in channel estimation compared with that of the rectangular window.
Simulation results verify the accuracy of the obtained analytical results and confirm the superiority of the proposed window designs in improving the channel estimation performance over the conventional rectangular or Sine windows.
\end{abstract}

\section{Introduction}

Future wireless networks are expected to provide high-speed and ultra-reliable communications for a wide range of emerging mobile applications\cite{Hadani2017orthogonal,XiaomingAccess,JiayiZhang5G,liu2019deep,liu2020deep,liu2019maximum,HaijunTeraHertzNOMA}, including online video gaming, unmanned aerial vehicles (UAV)\cite{HaijunZhang},  vehicle-to-everything (V2X), high-speed railway systems, etc.
In high-mobility channels, the multipath propagation and the temporal channel variations give rise to the frequency-selective fading (time dispersion) and time-selective fading (frequency dispersion), respectively, resulting in the so called \textit{doubly-selective} or \textit{doubly-dispersive} channels\cite{Ma2003Maximum}.
To cope with the channel dynamics, a new two-dimensional (2D) modulation scheme referred to as the orthogonal time-frequency space (OTFS) modulation was recently proposed in \cite{Hadani2017orthogonal} and has received an increasing amount of attention in both academia and industry, e.g. \cite{Hadani2017orthogonal,wei2020orthogonal,RavitejaOTFS,SurabhiDiversityAnalysisBER,Wei2021off}.

In OTFS modulation, data symbols are multiplexed in the delay-Doppler (DD) domain rather than in the time-frequency (TF) domain as in contrast to the traditional orthogonal frequency division multiplexing (OFDM) modulation\cite{RavitejaOTFS,WeiWindow,li2021cross}.
In practice, OTFS modulation effectively transforms the TF domain time-variant channel into an effective two-dimensional (2D) time-invariant channel in the DD domain, which exhibits both sparse and stable properties\cite{Hadani2017orthogonal,RavitejaOTFS}.
More importantly, the 2D transformation from the DD domain to the TF domain employed by an OTFS modulator allows the possibility of each information symbol to experience the whole TF domain channel over an OTFS frame.
Thus, OTFS enjoys the \textit{joint time-frequency diversity}\cite{SurabhiDiversityAnalysisBER} (the so-called \textit{full diversity} in \cite{Hadani2017orthogonal}), which is desirable to provide reliable communications over doubly dispersive channels.
In particular, it has been demonstrated that OTFS is resilient to severe delay-Doppler shifts and outperforms OFDM significantly for both uncoded \cite{RavitejaOTFS} and coded \cite{ZemenOrthogonalPrecoding,ShuangyangOTFS} systems.

{However, reliable communications with OTFS modulation highly rely on accurate channel estimation\cite{liu2020deep}, particularly in the presence of fractional Doppler\cite{RavitejaOTFS}, i.e., the exact Doppler frequency straddles a pair of finite-resolution bins rather than falls exactly into a bin in the Doppler domain.
Yet, most of existing works only considered integer Doppler for simplicity, e.g. \cite{FarhangCPOFDM,RavitejaPulseShapingFilterOTFS,ShenOTFSMassiveMIMOCE}.
In fact, ensuring integer Doppler requires a large speed separation among transceiver and all moving scatters to create a high Doppler resolution, which is not always possible in practical systems.
Although channel acquisition in the DD domain may be deemed more convenient than that in the TF domain in high mobility scenarios\cite{RavitejaOTFSCE,ShenOTFSMassiveMIMOCE}, the effective channel is spread across all the Doppler bins due to fractional Doppler, which sacrifices the DD domain channel sparsity.
Moreover, the channel estimation performance of OTFS systems is mainly limited by the inter-Doppler interference (IDI), where a guard space is usually required to avoid the IDI between data and pilot symbols, either employing a single pilot symbol \cite{RavitejaOTFSCE} or a pilot sequence\cite{ShenOTFSMassiveMIMOCE}.
Even worse, the IDI between data and pilot symbols caused by fractional Doppler becomes more severe leading to an error floor in the effective channel estimation.
As a remedy, to lower the error floor, a much larger guard space inserting between the data and pilot symbols is required, which causes a higher amount of signaling overhead.
Therefore, a pragmatic approach for reducing the channel spreading caused by fractional Doppler is desired.

As mentioned in \cite{FarhangCPOFDM,RezazadehReyhaniCPOFDM}, windowing in the TF domain has the potential in combating the effective channel spreading in the DD domain.
Yet, the authors in \cite{FarhangCPOFDM,RezazadehReyhaniCPOFDM} did not propose any method for window designs.
Moreover, the role of windowing in OTFS modulation and its impact on the performance of OTFS channel estimation are not well understood yet.
In the literature, to the best of our knowledge, there is no existing work studying on the window design for OTFS, which motivates this work.}

In this paper, we study the window design for OTFS modulation to improve the channel estimation performance with the consideration of practical fractional Doppler.
Firstly, the roles of windowing on OTFS systems, such as the effective channel and the corresponding  sparsity, are identified.
Secondly, we analyze the impact of windowing on the effective channel estimation performance.
Thirdly, we propose to employ a Dolph-Chebyshev (DC) window in the TF domain to facilitate the channel estimation in the DD domain.
The employed DC window is optimal in the sense that it can obtain a predefined channel sparsity while suppressing the channel spreading caused by the fractional Doppler to the largest degree.
Due to the enhanced channel sparsity, applying the proposed DC window at either the transmitter or the receiver can achieve a much lower channel estimation error floor compared with the conventional rectangular window.
Extensive simulations are conducted verify the analytical results and to demonstrate the substantial performance gain of the proposed window design over the conventional rectangular and Sine windows.

{\textit{Notations:} $\mathbb{Z}^{+}$ denotes the set of all non-negative integers;
	$\mathbb{C}^{M\times N}$ denotes the set of all $M\times N$ matrices with complex entries;
	$\abs{\cdot}$ denotes the absolute value of a complex scalar;
	$E\{\cdot\}$ denotes the expectation;
	$\left(\cdot\right)^*$ denotes the conjugate operation;
	$\left(\cdot\right)_N$ denotes the modulus operation with respect to $N$;
	$\Re\{\cdot\}$ returns the real part of the input complex number;
	$\lfloor\cdot\rfloor$ is the floor function which returns the largest integer smaller than the input value;
	The circularly symmetric complex Gaussian distribution with mean $\boldsymbol{\mu}$ and covariance matrix $\boldsymbol{\Sigma}$ is denoted by ${\cal CN}(\boldsymbol{\mu},\boldsymbol{\Sigma})$;
	$\sim$ stands for ``distributed as''.}

\begin{figure}[t]
	\centering
	\includegraphics[width=3.3in]{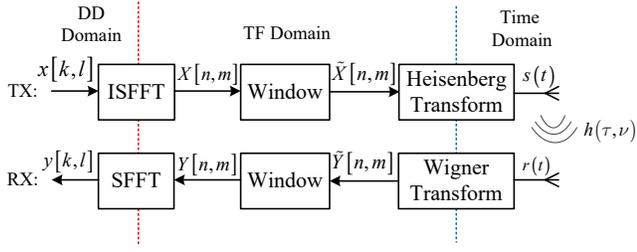}\vspace{-3mm}
	\caption{The block diagram of the OTFS transceiver\cite{RavitejaOTFS}.}\vspace{-5mm}
	\label{OFDM_OTFS}
\end{figure}

\section{System Model}

\subsection{OTFS Transmitter}

A practical implementation of the OTFS transceiver is shown in Fig. \ref{OFDM_OTFS}.
{Without loss of generality, we assume that one OTFS frame occupies a bandwidth of $B_{\mathrm{OTFS}}$ and a time duration of $T_{\mathrm{OTFS}}$.
The total available bandwidth $B_{\mathrm{OTFS}}$ is divided into $M$ subcarriers with an equal spacing of $\Delta f = \frac{B_{\mathrm{OTFS}}}{M}$.
The total time duration $T_{\mathrm{OTFS}}$ is divided into $N$ time slots with an equal-length slot duration of $T = \frac{T_{\mathrm{OTFS}}}{N}$.
As a result, a grid of $N\times M$ can be constructed in the TF domain.
Note that the delay resolution is determined by the reciprocal of the system bandwidth, i.e., $\frac{1}{M \Delta f}$, while the Doppler resolution is determined by the OTFS frame duration, i.e., $\frac{1}{NT}$ \cite{RavitejaOTFS}.
Correspondingly, in the DD domain, $N$ denotes the number of Doppler indices with a Doppler resolution of $\frac{1}{NT}$ and $M$ denotes the number of delay indices with a delay resolution of $\frac{1}{M \Delta f}$.}
Consider a baseband modulated symbol in the DD domain:
\begin{equation} \label{DD_DomainSymbol}
x\left[ {k,l} \right] \in \mathbb{A} = \{a_1,\ldots,a_Q\},
\end{equation}
where $k \in \{0,\ldots,N-1\}$ represents the Doppler index, $l \in \{0,\ldots,M-1\}$ represents the delay index, and $\mathbb{A}$ denotes the constellation set with a size of $Q$.
We assume that a normalized constellation is adopted, i.e., $E\left\{ {{{\left| {x\left[ {k,l} \right]} \right|}^2}} \right\} = 1$, and a proper scrambler is applied to scramble the output of the encoder such that it is reasonable to assume $E\left\{ {{{\left| {x\left[ {k,l} \right]} \right|}}} {{{\left| {x\left[ {k',l'} \right]} \right|}}}\right\} = 0$, $\forall k\neq k'$, $\forall l\neq l'$.
OTFS modulator performs a 2D transformation which maps the data symbols $x\left[ {k,l} \right]$ in the DD domain to $X\left[ {n,m} \right]$ in the TF domain.
In particular, such mapping can be realized by the inverse symplectic finite Fourier transform (ISFFT)\cite{Hadani2017orthogonal}:
\begin{equation} \label{OTFS_Mod}
X\left[ {n,m} \right] = \frac{1}{{\sqrt {NM} }}\sum\nolimits_{k = 0}^{N - 1} {\sum\nolimits_{l = 0}^{M - 1} {x\left[ {k,l} \right]{e^{j2\pi \left( {\frac{{nk}}{N} - \frac{{ml}}{M}} \right)}}} },
\end{equation}
where $n \in \{0,\ldots,N-1\}$ is the time slot index and $m \in \{0,\ldots,M-1\}$ is the subcarrier index.

A TF domain transmitter (TX) window $U\left[ {n,m} \right]$ is imposed through a point-wise multiplication with the TF domain signal $X\left[ {n,m} \right]$:
\begin{equation} \label{TF_WindowTx}
\widetilde{X}\left[ {n,m} \right] = U\left[ {n,m} \right]X\left[ {n,m} \right],
\end{equation}
where $U\left[ {n,m} \right] \in \mathbb{C}$ denotes the complex-valued TX window weighted on the point of $\left[ {n,m} \right]$ in the TF domain grid.
Then, a multicarrier modulator is adopted to transform the TF domain signal $\widetilde{X}\left[ {n,m} \right]$ to a time-domain signal $s\left( t \right)$, given by
\begin{equation} \label{MulticarrierModI}
s\left( t \right) = \sum\limits_{n = 0}^{N - 1} {\sum\limits_{m = 0}^{M - 1} {\widetilde X\left[ {n,m} \right]{{g_{{\rm{tx}}}}\left( {t - nT} \right){e^{j2\pi m\Delta f\left( {t - nT} \right)}}} }},
\end{equation}
which is referred to as the Heisenberg transform in\cite{Hadani2017orthogonal}, where $t$ denotes the continuous time variable.
The time domain function ${{g_{{\rm{tx}}}}\left( {t} \right)}$ is the pulse-shaping filter of the multicarrier modulator for the windowed TF domain symbol.

\subsection{DD Domain Channel Response}
For a linear time-variant channel, the received signal in the time domain is given by\cite{RavitejaOTFS}
\begin{equation}\label{LTVChannelII}
r\left( t \right) = \int {\int {h\left( {\tau ,\nu } \right)} } {{e^{j2\pi \nu \left( {t - \tau } \right)}}}s\left( {t - \tau } \right)d\tau d\nu + w\left(t\right),
\end{equation}
where $w\left(t\right)$ denotes the noise signal in the time domain following a stationary Gaussian random process and we have $w\left(t\right) \sim \mathcal{CN}\left(0,N_0\right)$ with $N_0$ denoting the noise variance.
In practice, only few reflectors are moving within one OTFS frame duration and thus only a small number of channel taps are associated with Doppler shift\cite{Hadani2017orthogonal,RavitejaOTFS}.
Therefore, the resulting channel response in the DD domain is sparse compared with the whole DD domain grid spanned by one OTFS frame.
In particular, considering a channel consisting of $P$ independent distinguishable paths, the channel response in the DD domain can be modeled by
\begin{equation}\label{LTVChannelDD}
{h\left( {\tau ,\nu } \right)} = \sum\nolimits_{i=1}^{P} {h_i} \delta(\tau-\tau_i)\delta(\nu-\nu_i),
\end{equation}
where $h_i \in \mathbb{C}$, $\tau_i$, and $\nu_i$ denote the channel coefficient, delay, and Doppler shift associated with the $i$-th path, respectively.
The variables $\tau_i$ and $\nu_i$ are defined as
$\tau_i = l_{\tau_i}\frac{1}{M\Delta f}$ and
$\nu_i = \left(k_{\nu_i} + \kappa_{\nu_i}\right)\frac{1}{N T}$, 
respectively, where $l_{\tau_i} \in \{0 ,\ldots, l_{\mathrm{max}}\}$, ${k_{{\nu _i}}} \in \{-k_{\mathrm{max}}, \ldots, k_{\mathrm{max}}\}$, and $-\frac{1}{2} < \kappa_{\nu_i} < \frac{1}{2}$ denote the integer delay, integer Doppler, and fractional Doppler indices, respectively.
Variables $k_{\max } \in \mathbb{Z}^+$ and $l_{\max } \in \mathbb{Z}^+$ denote the maximum Doppler and delay indices, respectively.

\subsection{OTFS Receiver}
At the receiver side, we first perform a multicarrier demodulation for the received signal $r\left( t \right)$ with a receiving filter to obtain the TF domain signal $\widetilde{Y}\left[n,m\right]$, given by:
\begin{equation}
\widetilde{Y}\left[n,m\right] = \int {r\left( t \right)g_{{\rm{rx}}}^ * \left( {t - nT} \right){e^{ - j2\pi m\Delta f\left( {t - nT} \right)}}dt},\label{MulticarrierDeMod}
\end{equation}
which is referred to as the Wigner transform in \cite{Hadani2017orthogonal}.
A time domain function ${{g_{{\rm{rx}}}}\left( {t} \right)}$ serving as a receiving filter for the multicarrier demodulator is adopted to sample the discrete symbol $\widetilde Y\left[ {n,m} \right]$ from the received waveform $r\left( t \right)$.
Substituting \eqref{MulticarrierModI}, \eqref{LTVChannelII}, and \eqref{LTVChannelDD} into \eqref{MulticarrierDeMod}, and assuming ideal transceiver pulse shaping filters satisfying the bi-orthogonal condition\cite{RavitejaOTFS}, we have
\begin{equation} \label{TFIOIdealPulse}
\widetilde{Y}\left(n,m\right) = {\widetilde X\left[ {n,m} \right]} \widetilde H\left[n,m\right]  + \widetilde{Z}\left[n,m\right],
\end{equation}
where the TF domain effective channel is given by
\begin{equation} \label{TFIOChannelIdealPulse}
\widetilde H\left[n,m\right] =\sum\limits_{i = 1}^{P}  {h_i {e^{ - j2\pi \frac{\left(k_{\nu_i} + \kappa_{\nu_i}\right)l_{\tau_i}}{NM}}} {e^{j2\pi \left( {\frac{{n\left(k_{\nu_i} + \kappa_{\nu_i}\right)}}{N} - \frac{{ml_{\tau_i}}}{M}} \right)}}  } .
\end{equation}

Corresponding to the TX window, we can insert a receiver (RX) window $V\left[ {n,m} \right]$ to the received signal in the TF domain:
\begin{equation} 
{Y}\left[ {n,m} \right] = V\left[ {n,m} \right]\widetilde{Y}\left[ {n,m} \right],\label{TF_WindowRx}
\end{equation}
where $V\left[ {n,m} \right] \in \mathbb{C}$.
Then, an OTFS demodulator transforms the TF domain signals ${Y}\left[ {n,m} \right]$ to the DD domain signals $y\left[ {k,l} \right]$ through a symplectic finite Fourier transform (SFFT) \cite{Hadani2017orthogonal}:
\begin{equation} \label{OTFS_DeMod}
y\left[ {k,l} \right] \hspace{-0.3mm}=\hspace{-0.3mm} \frac{1}{{\sqrt {NM} }}\sum\nolimits_{n = 0}^{N - 1} {\sum\nolimits_{m = 0}^{M - 1} {Y\left[ {n,m} \right]{e^{-j2\pi \left( {\frac{{kn}}{N} - \frac{{lm}}{M}} \right)}}} }.
\end{equation}

\section{The Impact of Windowing on Effective Channel}
In this section, we analyze the impacts of windowing on the effective channel for OTFS modulation.

\subsection{Effective Channel in the DD Domain}
According to the OTFS transceiver structure introduced above, the output of the OTFS demodulator in the DD domain is given by\cite{RavitejaOTFS}
\begin{align} \label{OTFS_DeModIdealPulse}
y\left[ {k,l} \right] 
&=  \sum\nolimits_{k' = 0}^{N - 1} {\sum\nolimits_{l' = 0}^{M - 1} {x\left[ {k',l'} \right]} } {h_w}\left[ {k - k',l - l'} \right] \notag\\
&+\sum\nolimits_{k' = 0}^{N - 1} {\sum\nolimits_{l' = 0}^{M - 1} {z\left[ {k',l'} \right]} } {v_z}\left[ {k - k',l - l'} \right].
\end{align}
In \eqref{OTFS_DeModIdealPulse}, ${h_w}\left[ {k ,l } \right]$ denotes the \textit{effective channel} in the DD domain capturing the windows' effect and it is given by
\begin{equation} \label{DDIOChannelDiscreteIdealPulseFractional}
{h_w}\left[ {k ,l } \right]
= \sum\limits_{i=1}^{P} \hspace{-0.3mm}{h_i} w(k\hspace{-0.3mm}-\hspace{-0.3mm}k_{\nu_i}\hspace{-0.3mm}-\hspace{-0.3mm}\kappa_{\nu_i}, l\hspace{-0.3mm}-\hspace{-0.3mm}l_{\tau_i}) {e^{ - j2\pi \frac{\left(k_{\nu_i} + \kappa_{\nu_i}\right)l_{\tau_i}}{NM} }},
\end{equation}
where $w(k - {k_{{\nu _i}}}-\kappa_{\nu_i},l - {l_{{\tau _i}}})$ is an equivalent DD domain filter designed by the TX-RX window which is given by\cite{RavitejaOTFS}
\begin{align} \label{DDFilterIdealPulse}
w(k - {k_{{\nu _i}}}-\kappa_{\nu_i},l - {l_{{\tau _i}}}) &= \frac{1}{NM}\sum\limits_{n = 0}^{N - 1} \sum\limits_{m = 0}^{M - 1} V\left[ {n,m} \right]U\left[ {n,m} \right] \notag\\[-1mm]
&\hspace{-5mm}\times {e^{ - j2\pi n\frac{\left( {k-k_{\nu_i} - \kappa_{\nu_i}} \right)}{N}}}{e^{j2\pi m \frac{\left( l-l_{\tau_i} \right)}{M }}}.
\end{align}
Also in \eqref{OTFS_DeModIdealPulse}, ${v_z}\left[ {k ,l } \right]$ is a DD domain filter induced only by the RX window and is given by
\begin{equation} \label{DDFilterRxIdealPulse}
{v_z}\left[ {k ,l } \right] \hspace{-0.1mm}= \hspace{-0.1mm} \frac{1}{NM}\sum\nolimits_{n = 0}^{N - 1} \sum\nolimits_{m = 0}^{M - 1} V\left[ {n,m} \right]{e^{ - j2\pi \frac{nk}{N}}}{e^{j2\pi \frac{ml}{M}}}.
\end{equation}

We can observe that different from the original DD domain channel response in \eqref{LTVChannelDD}, the effective channel in \eqref{DDIOChannelDiscreteIdealPulseFractional} has a circular structure due to ${h_w}\left[ {\left(k\right)_N ,\left(l\right)_M } \right] = {h_w}\left[ {k ,l } \right]$.
As such, from \eqref{OTFS_DeModIdealPulse}, we can observe that the received signal $y\left[ {k,l} \right]$ is a 2D circular convolution between the data symbols, ${x\left[ {k,l} \right]}$, and the effective channel, ${h_w}\left[ {k ,l } \right]$, in the DD domain.
Furthermore, as the data and training symbols are multiplexed in the DD domain \cite{RavitejaOTFSCE}, the channel estimation performance and the data detection complexity depend on the effective channel ${h_w}\left[ {k ,l } \right]$ instead of the original channel response $h\left( {\tau ,\nu } \right)$.
As shown in \eqref{DDIOChannelDiscreteIdealPulseFractional}, the effective channel ${h_w}\left[ {k ,l } \right]$ is a summation of the channel spread of each path where $w(k - {k_{{\nu _i}}}-\kappa_{\nu_i},l - {l_{{\tau _i}}}) \neq 0$, $\forall i$, and the spreading pattern can be manipulated by the design of the DD domain filter $w(k,l)$.
In other words, the channel sparsity of the effective channels can be controlled by the TX and RX windows.
In \eqref{DDFilterIdealPulse}, we can observe that imposing a TX window $U\left[ {n,m} \right]$ or a RX window $V\left[ {n,m} \right]$ in the TF domain has the same effect on the design of the DD domain filter $w(k,l)$.
In contrast, only the RX window $V\left[ {n,m} \right]$ affects the DD domain filter, ${v_z}\left[ {k,l} \right] $, which alters the properties of the noise at the receiver side.

\subsection{Effective Channel Sparsity with Rectangular Window}
Since the rectangular window, i.e., $V\left[ {n,m} \right] = U\left[ {n,m} \right] =1$, $\forall n,m$, is the most straightforward one to be considered\cite{RavitejaOTFS,RavitejaOTFSCE}, we investigate the  effective channel sparsity with rectangular window for both cases of integer and fractional Doppler\footnote{Different from \cite{RavitejaOTFS,RavitejaOTFSCE}, we focus on discussing the reduced effective channel sparsity due to the existence of fractional Doppler, which motivates our analysis and design in the following sections.}.
With employing the rectangular window, we have the DD domain filters given by\cite{RavitejaOTFS}
\begin{align}
w(k \hspace{-0.5mm}-\hspace{-0.5mm} {k_{{\nu _i}}}\hspace{-0.5mm}-\hspace{-0.5mm}\kappa_{\nu_i},l \hspace{-0.5mm}-\hspace{-0.5mm} {l_{{\tau _i}}}) &= \mathcal{G}^{\mathrm{Rect}}_N\left({k \hspace{-0.5mm}-\hspace{-0.5mm} {k_{{\nu _i}}} \hspace{-0.5mm} -\hspace{-0.5mm}\kappa_{\nu_i}}\right) \mathcal{F}^{\mathrm{Rect}}_M \left({l\hspace{-0.5mm} -\hspace{-0.5mm} {l_{{\tau _i}}}}\right)\notag\\[-0.5mm]
\text{and}\; {v_z}\left[ {k ,l } \right] &= \mathcal{G}^{\mathrm{Rect}}_N\left(k\right) \mathcal{F}^{\mathrm{Rect}}_M \left(l\right),\label{WindowDDDomainFractionalNoise}
\end{align}
respectively.
Functions $\mathcal{G}^{\mathrm{Rect}}_N\left(k\right)$ and $\mathcal{F}^{\mathrm{Rect}}_M \left(l\right)$ represent the filters in the delay and Doppler domains, respectively, and they are given by $\mathcal{G}^{\mathrm{Rect}}_N\left(k\right) = \frac{1}{N}\left({e^{ - j\left( {N - 1} \right) \frac{\pi k}{N}}}\frac{\sin \left(\pi k\right)}{{\sin \left( {\frac{{\pi k }}{N}} \right)}} \right)$ and $\mathcal{F}^{\mathrm{Rect}}_M \left(l\right) = \frac{1}{M}\left({{e^{ - j\left( {M - 1} \right) \frac{\pi l}{M}}}\frac{{\sin \left( {\pi l } \right)}}{{\sin \left( {\frac{{\pi l}}{M}} \right)}}} \right)$, respectively.

For the case of integer Doppler, i.e., $\kappa_{\nu_i} = 0$, the DD domain filter is simplified as
\begin{align}\label{WindowDDDomain}
w\left[k - {k_{{\nu _i}}},l - {l_{{\tau _i}}}\right] &= \delta\left[k - {k_{{\nu _i}}}\right] \delta\left[l - {l_{{\tau _i}}}\right] \\
&\hspace{-10mm}= \left\{ {\begin{array}{*{20}{c}}
	{1}&{\left(k - {k_{{\nu _i}}}\right)_N = 0,\left(l - {l_{{\tau _i}}}\right)_M = 0}\\
	0& \mathrm{otherwise}
	\end{array}} \right.,\notag
\end{align}
and the effective channel in the DD domain is given by
\begin{equation} \label{DDIOChannelDiscreteIdealPulseInteger}
{h_w}\left[ {k ,l } \right]
= \sum\nolimits_{i=1}^{P} {h_i} \delta\left[k - {k_{{\nu _i}}}\right] \delta\left[l - {l_{{\tau _i}}}\right] {e^{ - j2\pi \frac{k_{\nu_i}l_{\tau_i}}{NM} }}.
\end{equation}
We can observe that the effective channel ${h_w}\left[ {k ,l } \right]$ in the DD domain has a response if and only if $k = {k_{{\nu _i}}}$ and $l = {l_{{\tau _i}}}$, i.e., the effective channel shares the same channel sparsity with the original DD domain channel response in \eqref{LTVChannelDD}.
Moreover, the effective channel ${h_w}\left[ {k ,l } \right]$ is a phase-rotated version of the original channel response in \eqref{LTVChannelDD}, where the delay and Doppler shift of the $i$-th path rotates the original channel response ${h_i}$ with a phase of $2\pi k_{\nu_i}l_{\tau_i} /MN$.

\begin{figure}[t]
	\centering
	\includegraphics[width=3.5in]{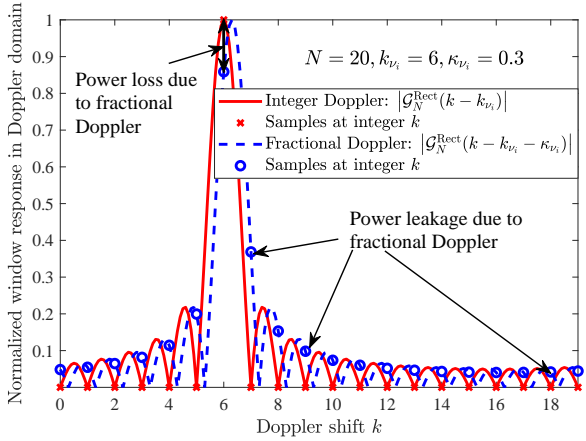}\vspace{-3mm}
	\caption{The channel spreading in the Doppler domain with a rectangular window with/without fractional Doppler, where $\mathrm{SL}_{w} \approx \frac{1}{N}$ denotes the sidelobe level of the adopted rectangular window.}\vspace{-5mm}
	\label{FractionalIDI}%
\end{figure}

For the case of fractional Doppler, i.e., $\kappa_{\nu_i} \neq 0$, the effective channel is given by
\begin{equation} \label{DDIOChannelDiscreteIdealPulseFractional_RectWindow}
{h_w}\hspace{-0.5mm}\left[ {k ,l } \right] \hspace{-0.5mm}
= \hspace{-0.5mm}\sum\limits_{i=1}^{P} {h_i} \mathcal{G}^{\mathrm{Rect}}_N\hspace{-0.5mm}\left({k \hspace{-0.5mm}-\hspace{-0.5mm} {k_{{\nu _i}}}  \hspace{-0.5mm}-\hspace{-0.5mm}\kappa_{\nu_i}}\right)\hspace{-0.5mm} \delta\hspace{-0.5mm}\left[l\hspace{-0.5mm} -\hspace{-0.5mm} {l_{{\tau _i}}}\right] 
 {e^{ - j2\pi \frac{\left(k_{\nu_i} \hspace{-0.25mm}+\hspace{-0.25mm} \kappa_{\nu_i}\right)l_{\tau_i}}{NM} }}. 
\end{equation}
From \eqref{DDIOChannelDiscreteIdealPulseFractional_RectWindow}, we can observe that the effective channel in the DD domain ${h_w}\left[ {k ,l } \right]$ contains more ``paths'' (non-zero entries) than that of the original channel response in \eqref{LTVChannelDD}, since the Doppler domain filter $\mathcal{G}^{\mathrm{Rect}}_N\left({k - {k_{{\nu _i}}}  -\kappa_{\nu_i}}\right) \neq 0$, $\forall k,{k_{{\nu _i}}}$, and $\forall \kappa_{\nu_i} \neq 0$.
In fact, for each path with a Doppler shift of ${k_{{\nu _i}}} + \kappa_{\nu_i}$, the channel coefficient ${h_i}$ is spread to all the Doppler indices $k$ in the Doppler domain.
To visualize the channel spreading, we ignore the delay domain at the moment and plot the Doppler domain filter response $\left|\mathcal{G}^{\mathrm{Rect}}_N\left({k - {k_{{\nu _i}}}  -\kappa_{\nu_i}}\right)\right|$ to illustrate the impact of  fractional Doppler in Fig. \ref{FractionalIDI}.
It can be seen that without the fractional Doppler, the filter $\left|\mathcal{G}^{\mathrm{Rect}}_N\left({k - {k_{{\nu _i}}}}\right)\right|$ is a perfect sampling function $\delta \left[k - {k_{{\nu _i}}}\right]$, i.e., no channel spread.
However, the existence of the fractional Doppler shift $\kappa_{\nu_i}$ not only reduces signal power at the sampling point $k = {k_{{\nu _i}}}$, but also introduces non-negligible power leakage from the Doppler shift ${k_{{\nu _i}}}$ to $k \neq  {k_{{\nu _i}}}$.
In other words, with the application of the rectangular window, fractional Doppler sacrifices the sparsity of the effective channel in the DD domain, which could degrade the channel estimation performance and increase the complexity of data detection.
Therefore, it is desired to design a window which can null/suppress the power leakage and improve the effective channel sparsity.

\begin{figure*}[!t]
	\setcounter{equation}{23}
	\begin{align}\label{InterferenceLevel}
	E\left\{ {{{\left| {I\left[ {k,l} \right]} \right|}^2}} \right\} 
	&= \sum\limits_{k' \notin \mathcal{K}} {\sum\limits_{l' = l-l_{\mathrm{max}}}^{{l }} {E\left\{ {{{\left| {x\left[ {k',l'} \right]} \right|}^2}} \right\} E\left\{ {{{\left| {{h_w}\left[ {{{\left( {k - k'} \right)}_N},{{\left( {l - l'} \right)}_M}} \right]} \right|}^2}} \right\}} }\notag\\[-1mm]
	& \mathop  = \limits^{(a)}  \sum\limits_{k' \notin \mathcal{K}} \sum\limits_{ l-l' \in [0,l_{\mathrm{max}}], {{\left( {l - l'} \right)}_M} = l_{\tau_i}} {{E\left\{ {{{\left| {\sum\limits_{i = 1}^P {{h_i}} w\left( {{{\left( {k - k'} \right)}_N} - {k_{{\nu _i}}} - {\kappa _{{\nu _i}}},0} \right){e^{ - j2\pi \frac{{\left( {{k_{{\nu _i}}} + {\kappa _{{\nu _i}}}} \right){l_{{\tau _i}}}}}{{NM}}}}} \right|}^2}} \right\}} }
	\end{align}
	\hrulefill
	\vspace{-5mm}
\end{figure*}

\begin{figure*}
	\setcounter{equation}{26}
	\begin{align}
	\mathrm{MSE} &= \sum_{k = {k_p} - {k_{\max }} - \hat k}^{{k_p} + {k_{\max }} + \hat k} \sum_{l = {l_p}}^{{l_p} + {l_{\mathrm{max}} }} \hspace{-0.5mm} E\hspace{-0.5mm}\left\{ \hspace{-0.5mm} {{{\left| {h_w}\hspace{-0.5mm}\left[\hspace{-0.5mm} {\left(k\hspace{-0.5mm}-\hspace{-0.5mm}k_p\right)_N,\left(l\hspace{-0.5mm}-\hspace{-0.5mm}l_p\right)_M } \hspace{-0.3mm} \right] \hspace{-0.5mm}-\hspace{-0.5mm} {\hat{h}_w}\left[\hspace{-0.5mm} {\left(k\hspace{-0.5mm}-\hspace{-0.5mm}k_p\right)_N ,\left(l\hspace{-0.5mm}-\hspace{-0.5mm}l_p\right)_M }\hspace{-0.3mm} \right] \right|}^2}} \hspace{-0.5mm} \right\} \label{ChannelESTIMATIONMSE_I} \\[-1mm]
	\lim\limits_{N_0 \to 0}\mathrm{MSE} &= \sum_{k = {k_p} - {k_{\max }} - \hat k}^{{k_p} + {k_{\max }} + \hat k} \sum_{l = {l_p}}^{{l_p} + {l_{\mathrm{max}} }} \frac{E\left\{ {{{\left| {I\left[ {k,l} \right]} \right|}^2}} \right\}}{\left|x_p\right|^2} \approx \left(N\hspace{-0.5mm}-\hspace{-0.5mm}4{k_{\max }} \hspace{-0.5mm}-\hspace{-0.5mm} 4\hat k \hspace{-0.5mm}-\hspace{-0.5mm} 1\right) \left(2{k_{\max }} \hspace{-0.5mm}+\hspace{-0.5mm} 2\hat k\hspace{-0.5mm}+\hspace{-0.5mm}1\right) \left(l_{\mathrm{max}} \hspace{-0.5mm}+\hspace{-0.5mm} 1\right)\mathrm{SL}_{w}^2  \label{ChannelESTIMATIONMSE_iI} 
	\end{align}
	\hrulefill
	\vspace{-5mm}
	\setcounter{equation}{19}
\end{figure*}

\section{Effective Channel Estimation Performance with Arbitrary Window}
In this work, we adopt the channel estimation scheme proposed in \cite{RavitejaOTFSCE}, where a single pilot symbol is embedded in the DD domain and a guard space is inserted between the pilot symbol and data symbols.
In fact, to the best of our knowledge, the channel estimation scheme in \cite{RavitejaOTFSCE} is the first DD domain channel estimation method proposed for OTFS in the literature, which is simple and practical.
In this section, we investigate the impact of windowing on channel estimation performance based on the scheme in \cite{RavitejaOTFSCE}.
Let us assume that the only pilot symbol $x_p$ is inserted at the $\left[k_p,l_p\right]$-th DD grid and data symbols ${x_d}\left[ {k,l} \right]$ are arranged as follow\cite{RavitejaOTFSCE}
\begin{equation}\label{PilotEmbedded}
x\left[ {k,l} \right] = \left\{ {\begin{array}{*{20}{c}}
	{{x_p}}&{k = {k_p},l = {l_p}},\\
		0&\begin{array}{l}
	k \in \mathcal{K}, k \neq {k_p}, l \in \mathcal{L}, l \neq {l_p},
	\end{array}\\
	{{x_d}\left[ {k,l} \right]}&{{\rm{otherwise}}},
	\end{array}} \right.
\end{equation}
where $\mathcal{K} = \{{k_p} - 2{k_{\max }} - 2\hat k, \ldots, {k_p} + 2{k_{\max }} + 2\hat k\}$ and $\mathcal{L} = \{{l_p} - {l_{\mathrm{max}} }, \ldots, {l_p} + {l_{\mathrm{max}} }\}$ denotes the index sets of the guard space in the Doppler and delay domains, respectively.
Variable $\hat k \in \mathbb{Z}^+$ denotes the additional guard to mitigate the spread due to fractional Doppler and $\hat k \in \left\{0,\ldots,  \lfloor\frac{N-4k_{\max }-1}{4}\rfloor \right\}$.
Increasing $\hat k$ would potentially increase the channel estimation performance while reduces the spectral efficiency, as the signaling overhead increases with $\hat k$, i.e., the total signaling overhead is $\left(2{l_{\mathrm{max}} } +1\right)\left(4{k_{\max }} + 4\hat k + 1\right)$.

The estimation of the effective channel is based on the received signals in the DD domain, which are given by
\begin{align}\label{EEChannelEstimation}
y\left[k,l\right] &= x_p {h_w}\left[ {\left(k-k_p\right)_N ,\left(l-l_p\right)_M } \right] + I\left[k,l\right] \notag\\
&+ \sum\nolimits_{k' = 0}^{N - 1} {\sum\nolimits_{l' = 0}^{M - 1} {z\left[ {k',l'} \right]} } {v_z}\left[ {k - k',l - l'} \right],
\end{align}
where ${k_p} - {k_{\max }} - \hat k \le k \le {k_p} + {k_{\max }} + \hat k$ and ${l_p} \le l \le {l_p} + {l_{\mathrm{max}} }$.
According to \cite{RavitejaOTFSCE}, the effective DD domain channel can be estimated by
\begin{equation}\label{DDCEFormula}
	{\hat h_w}\left[ {\left(k\hspace{-0.5mm}-\hspace{-0.5mm}k_p\right)_N ,\left(l\hspace{-0.5mm}-\hspace{-0.5mm}l_p\right)_M } \right] = \frac{y\left[k,l\right]}{x_p},\; \text{if} \left|y\left[k,l\right]\right| \ge 3\sqrt{N_0}.
\end{equation}
In \eqref{EEChannelEstimation}, $I\left[k,l\right]$ denotes the interference spread from data symbols due to the existence of fractional Doppler, which is given by
\begin{equation}\label{InterferenceCE}
I\left[k,l\right] = \sum_{k' \notin \mathcal{K}} \sum_{l'=0}^{l_{\mathrm{max}}} x\left[ {k',\left(l-l'\right)_M} \right] {h_w}\left[ {\left(k-k'\right)_N ,l' } \right].
\end{equation}
We can observe that in the delay domain, only $l_{\mathrm{max}}+1$ symbols before $l$ affect the received symbol on $l$.
On the other hand, in the Doppler domain, all the data symbols outside the guard space $k' \notin \mathcal{K}$ affect the received symbol on $k$.
Due to the existence of the interference term $I\left[k,l\right]$, the channel estimation in \eqref{DDCEFormula} suffers from an error floor even increasing the system signal-to-noise ratio (SNR).
Note that when applying the full guard space\cite{RavitejaOTFSCE}, i.e., $4{k_{\max }} + 4\hat k + 1 = N$, the interference term in \eqref{InterferenceCE} would disappear and there is no error floor in the effective channel estimation.
However, it requires a higher signaling overhead of $\left(2{l_{\mathrm{max}} } +1\right)N$ compared with that of the scheme in \eqref{PilotEmbedded}.

In the following, we derive the interference power to investigate the impact of windowing on the effective channel estimation performance.
Since the transmitted data symbols are independent, the interference power can be calculated as \eqref{InterferenceLevel} {at the top of next page,} where the equality $(a)$ is obtained since only the data symbols on ${{\left( {l - l'} \right)}_M} = l_{\tau_i}$ in the summation over $l'$ affect the received symbol on $l$ with adopting a rectangular window in the delay domain, i.e., $V\left[ {n,m} \right] = V\left[ {n,m'} \right] = U\left[ {n,m} \right] = U\left[ {n,m'} \right]$, $\forall n,m,m'$, and $E\left\{ {{{\left| {x\left[ {k',l'} \right]} \right|}^2}} \right\} = 1$.
Assuming independent channel coefficients, i.e., $E\left\{{h_i}{h^*_j}\right\} = 0$, $\forall i \neq j$, \eqref{InterferenceLevel} becomes
\setcounter{equation}{24}
\begin{equation}
E\hspace{-0.5mm}\left\{\hspace{-0.5mm} {{{\left| {I\left[ {k,l} \right]} \right|}^2}} \hspace{-0.5mm}\right\} \hspace{-1mm}=\hspace{-1mm} \sum\limits_{k' \notin \mathcal{K}} \hspace{-0.5mm} \sum\limits_{i = 1}^P \hspace{-0.5mm} {E\hspace{-0.5mm}\left\{\hspace{-0.5mm} {{{\left| {{h_i}} \right|}^2}} \hspace{-0.5mm}\right\}} {{\left| {w\left( {{{\left( {k \hspace{-0.5mm}-\hspace{-0.5mm} k'} \right)}_N} \hspace{-0.5mm}-\hspace{-0.5mm} {k_{{\nu _i}}} \hspace{-0.5mm}-\hspace{-0.5mm} {\kappa _{{\nu _i}}},0} \right)} \right|}^2}.
\end{equation}

It can be observed that the interference power is determined by the window response at ${{\left( {k - k'} \right)}_N} - {k_{{\nu _i}}} - {\kappa _{{\nu _i}}}$.
Thanks to the guard space, the window response $\left| {w\left( {{{\left( {k - k'} \right)}_N} - {k_{{\nu _i}}} - {\kappa _{{\nu _i}}},0} \right)} \right|$ lies in its sidelobe and becomes almost a constant, as shown in Fig. \ref{FractionalIDI}.
Therefore, we assume $\left| {w\left( {{{\left( {k - k'} \right)}_N} - {k_{{\nu _i}}}- {\kappa _{{\nu _i}}},0} \right)} \right| \approx \mathrm{SL}_{w}$ for ${k' \notin \mathcal{K}}$ and ${k_p} - {k_{\max }} - \hat k \le k \le {k_p} + {k_{\max }} + \hat k$.
Considering a normalized channel power gain, i.e., $\sum\nolimits_{i = 1}^P {E\left\{ {{{\left| {{h_i}} \right|}^2}} \right\}} = 1$, the average interference power can be approximated by
\begin{equation}
E\left\{ {{{\left| {I\left[ {k,l} \right]} \right|}^2}} \right\}  \approx \left(N-4{k_{\max }} - 4\hat k - 1\right) \mathrm{SL}_{w}^2,
\end{equation}
where $\mathrm{SL}_{w}$ denotes the sidelobe level of the adopted window.
For instance, as shown in Fig. \ref{FractionalIDI}, we have $\mathrm{SL}_{w} \approx \frac{1}{N}$ for the case of rectangular window.
Define the mean squared error (MSE) of the effective channel estimation in the guard space as \eqref{ChannelESTIMATIONMSE_I} at the top of this page.
According to \eqref{EEChannelEstimation}, in high SNR regime, i.e., $N_0 \to 0$, the MSE of the effective channel estimation is given by \eqref{ChannelESTIMATIONMSE_iI} at the top of this page, which indicates the effective channel estimation error floor.
Note that the analytical result in \eqref{ChannelESTIMATIONMSE_iI} is applicable to arbitrary window.

Now, we can observe that in the high SNR regime, the error floor level in the effective channel estimation in \eqref{ChannelESTIMATIONMSE_iI} depends on the additional guard $\hat k$ and the sidelobe level $\mathrm{SL}_{w}$ of the designed window response.
{It can be seen that the MSE of the effective channel estimation is a quadratic function with respect to $\hat k$.
After some mathematical manipulations, it can be seen that when $\frac{N-8{k_{\max }-3}}{4} \le 0$, increasing $\hat k $ in the range of $\left\{0,\ldots,  \lfloor\frac{N-4k_{\max }-1}{4}\rfloor \right\}$ always results in a lower error floor level at the expense of more signaling overhead.
On the other hand, when $\frac{N-8{k_{\max }-3}}{4} \ge 1$, increasing $\hat k $ first increases and then decreases the MSE of effective channel estimation.
This is because for large $N$, increasing additional guard $\hat k $ introduces more entries to be estimated within the guard space, thereby might increasing channel estimation error.
Note that further increasing $\hat k $ reduces the IDI caused by the data symbols and thus reduces the effective channel estimation MSE, but it also consumes more signaling overhead.}
More importantly, as shown in \eqref{ChannelESTIMATIONMSE_iI}, a proper design of window response can achieve a low sidelobe level at the first place, which can effectively decrease the error floor level with a relatively small $\hat k$.
In fact, a window response with a low sidelobe level can enhance the effective channel sparsity, which can improve the channel estimation performance.
Moreover, as the TX and RX windows have the same impact on the effective channel in the DD domain in \eqref{DDIOChannelDiscreteIdealPulseFractional}, imposing a window at either the transmitter or the receiver will result in the same channel estimation error floor.

\section{Window Designs for OTFS Channel Estimation}
Based on the above discussed properties, we first discuss the ideal window response, which is not realizable but provides insightful guidelines for practical window designs.
Then, we propose to apply the DC window to enhance the effective channel sparsity, which will significantly improve the performance of both channel estimation and data detection.

\subsection{Ideal Window}
To facilitate the window design, we consider a separable TF domain window as follows:
\setcounter{equation}{28}
\begin{equation}
V\left[ {n,m} \right] \hspace{-0.5mm}=\hspace{-0.5mm} V_{\nu}\left[ {n} \right] V_{\tau}\left[ {m} \right]\;\text{and}\;
U\left[ {n,m} \right] \hspace{-0.5mm}=\hspace{-0.5mm} U_{\nu}\left[ {n} \right] U_{\tau}\left[ {m} \right],
\end{equation}
where $V_{\nu}\left[ {n} \right]$ and $U_{\nu}\left[ {n} \right]$ denote the RX and TX windows in the Doppler domain, respectively, and $V_{\tau}\left[ {m} \right]$ and $U_{\tau}\left[ {m} \right]$ denote the RX and TX windows in the delay domain, respectively.
As a result, the window response in the DD domain in \eqref{DDFilterIdealPulse} can be decomposed as
\begin{equation}\label{ArbitraryWindow}
w(k \hspace{-0.5mm} - \hspace{-0.5mm}{k_{{\nu _i}}}\hspace{-0.5mm}-\hspace{-0.5mm}\kappa_{\nu_i},l \hspace{-0.5mm}-\hspace{-0.5mm} {l_{{\tau _i}}}) = \mathcal{G}_N\left({k\hspace{-0.5mm} -\hspace{-0.5mm} {k_{{\nu _i}}} \hspace{-0.5mm} -\hspace{-0.5mm}\kappa_{\nu_i}}\right) \mathcal{F}_M \left({l \hspace{-0.5mm}-\hspace{-0.5mm} {l_{{\tau _i}}}}\right),
\end{equation}
where 
\begin{align}\label{WindowResponseFunctionDoppler}
	\hspace{-0.5mm}\mathcal{G}_N\left({k\hspace{-0.5mm} -\hspace{-0.5mm} {k_{{\nu _i}}} \hspace{-0.5mm} -\hspace{-0.5mm}\kappa_{\nu_i}}\right) \hspace{-0.5mm}&=\hspace{-0.5mm} \frac{1}{N}\hspace{-1mm}\sum\nolimits_{n = 0}^{N - 1}\hspace{-1mm}
	V_{\nu}\left[ {n} \right]U_{\nu}\left[ {n} \right]{e^{ - j2\pi n\frac{\left( {k\hspace{-0.25mm}-\hspace{-0.25mm}k_{\nu_i}\hspace{-0.25mm} -\hspace{-0.25mm} \kappa_{\nu_i}} \right)}{N}}} \notag\\[-0.5mm]
	\text{and}\;\mathcal{F}_M \left({l \hspace{-0.5mm}-\hspace{-0.5mm} {l_{{\tau _i}}}}\right) \hspace{-0.5mm}&=\hspace{-0.5mm} \frac{1}{M}\hspace{-1mm}\sum\nolimits_{m = 0}^{M - 1} \hspace{-1mm} V_{\tau}\left[ {m} \right]U_{\tau}\left[ {m} \right]{e^{j2\pi m \frac{\left( l\hspace{-0.25mm}-\hspace{-0.25mm}l_{\tau_i} \right)}{M}}}.
\end{align}

Combining \eqref{DDIOChannelDiscreteIdealPulseFractional} and \eqref{ArbitraryWindow}, the effective channel in the DD domain can be rewritten as
\begin{equation}
{h_w}\hspace{-0.5mm}\left[ {k ,l } \right]\hspace{-1mm}
=\hspace{-1mm}\sum\nolimits_{i = 1}^P \hspace{-0.5mm} {h_i} \mathcal{G}_N\hspace{-0.5mm}\left({k\hspace{-0.5mm} -\hspace{-0.5mm} {k_{{\nu _i}}} \hspace{-1mm} -\hspace{-0.5mm}\kappa_{\nu_i}}\right)\hspace{-0.5mm} \mathcal{F}_M \hspace{-0.5mm}\left({l \hspace{-0.5mm}-\hspace{-0.5mm} {l_{{\tau _i}}}}\right) \hspace{-0.5mm} {e^{ - j2\pi \frac{\left(k_{\nu_i} \hspace{-0.25mm}+ \hspace{-0.25mm}\kappa_{\nu_i}\right)l_{\tau_i}}{NM} }}.\label{OTFS_DeModIdealPulseIIIFractional}
\end{equation}
Since the delay resolution is usually sufficient and there is only negligible channel spread in the delay domain \cite{RavitejaOTFS}, the optimal window in the delay domain should be maintained as the rectangular window, i.e., $\mathcal{F}^{\mathrm{Ideal}}_M \left({l}\right) = \mathcal{F}^{\mathrm{Rect}}_M \left({l }\right)$.
Besides, with the existence of the fractional Doppler, as $-\frac{1}{2} < \kappa_{\nu_i} < \frac{1}{2}$, the ideal window in the Doppler domain is given by:
\begin{equation}\label{IdealWindow}
{{\cal G}^{\mathrm{Ideal}}_N}\left( k \right) = \left\{ {\begin{array}{*{20}{c}}
	{1,}&{ - 0.5 \le k \le 0.5,}\\[-0.5mm]
	{0,}&{{\rm{otherwise.}}}
	\end{array}} \right.
\end{equation}
In Fig. \ref{DifferentWindow}, we illustrate that the ideal window can tolerate the fractional Doppler shift without sacrificing any channel gain and causing any channel spread.
However, to implement such an ideal window response, an infinite length of window in the time domain is needed, i.e., $N \to \infty$.
Recall that the fractional Doppler is caused by the finite $N$.
Therefore, the ideal window in \eqref{IdealWindow} is not realizable in practice.

\subsection{Dolph-Chebyshev Window}
In what follows, we propose to apply the Dolph-Chebyshev (DC) window at the transmitter or the receiver to improve the channel sparsity when channel state information (CSI) is not available.
In fact, it has been proved that the DC window is effective \cite{Dolph1946} in the sense that: 1) given the specified sidelobe level, the width of the mainlobe in the window response is the narrowest; or 2) given the fixed mainlobe width, the sidelobe level is minimized.
Note that the effective channel only has a considerably large entry when it is located in the mainlobe of the window response function in \eqref{WindowResponseFunctionDoppler}.
Therefore, given any channel sparsity requirement, the channel spread to other Doppler indices is reduced to the largest degree by using a DC window.

\begin{figure}[t]
	\centering 
	\includegraphics[width=3.3in]{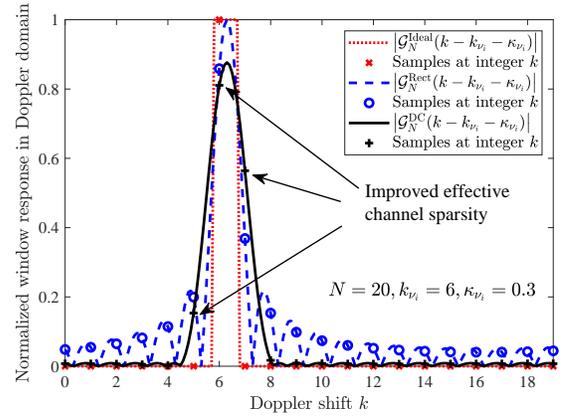}\vspace{-3mm}
	\caption{The effective channel in the Doppler domain with different types of windows and fractional Dopplers.}\vspace{-5mm}
	\label{DifferentWindow}%
\end{figure}

Particularly, if the required mainlobe width of the window response in the Doppler domain is $k_{\mathrm{main}} > 1$, the number of non-negligible effective channel spreading of each path in the Doppler domain is no more than $k_{\mathrm{main}}$, i.e., the effective channel sparsity is improved.
In this case, the lowest sidelobe level achieved by the DC window is \cite{WeiBeamWidthControl}
\begin{equation}
\mathrm{SL}_w [\mathrm{dB}] \hspace{-0.5mm}=\hspace{-0.5mm} -20 \log_{10} {{\cosh \left(\hspace{-0.5mm} {\frac{N}{2}{{\cosh }^{-1}}\hspace{-0.5mm}\left(\hspace{-0.5mm} \frac{3-\cos\left(\frac{k_{\mathrm{main}}}{2}\right)}{1+\cos\left(\frac{k_{\mathrm{main}}}{2}\right)} \hspace{-0.5mm}\right)}\hspace{-0.5mm} \right)}}.
\end{equation}
In this paper, to reveal the insights of employing TX/RX windows, we only adopt the DC window at the transmitter or the receiver side with the other side adopting a rectangular window. 
The TX window $U_{\nu}\left[ {n} \right]$ or the RX window $V_{\nu}\left[ {n} \right]$ can be obtained by (16) in \cite{Duhamel1953} according to the selected $\mathrm{SL}_w$.
In Fig. \ref{DifferentWindow}, for the same setting with Fig. \ref{FractionalIDI}, we employ a DC window at the transmitter side with $\mathrm{SL}_w [\mathrm{dB}]
 = -40$ dB and $k_{\mathrm{main}} \approx 3$.
We can observe that the resulting effective channel only has approximately $3$ entries with considerable gains and the channel spreading to all the other Doppler indices has been significantly suppressed due to the $40$ dB attenuation on the sidelobe  introduced by the designed DC window.
By putting $\mathrm{SL}_w [\mathrm{dB}]
= -40$ dB into \eqref{ChannelESTIMATIONMSE_iI}, we can find that the proposed DC window can effectively suppress the MSE of effective channel estimation,  compared with the rectangular window.

\section{Numerical Results}

In this section, we verify the accuracy of the derived analytical results and the effectiveness of the proposed designs via simulations.
For each OTFS frame, we set $N= 20$, $M = 30$, carrier frequency $f_c = 3$ GHz, and the subcarrier spacing is $\Delta f = 5$ kHz.
Without loss of generality, we set the maximum delay index as $l_{\mathrm{max}} = 4$ and the maximum Doppler index as $k_{\mathrm{max}} = 3$, corresponding to the relative speed between the transceiver as $270$ km/h.
The number of paths in the DD domain is $P = 5$ and the additional guard space is $\hat k = [0,1]$.
For each channel realization, we randomly select the delay and Doppler indices such that we have $-k_{\mathrm{max}} \le {k_{{\nu _i}}} \le k_{\mathrm{max}}$ and $0 \le l_{\tau_i} \le l_{\mathrm{max}}$.
The channel coefficients $h_i$ are generated according to the distribution $h_{i}\sim \mathcal{CN}(0,q^{l_{\tau_i}})$, where $q^{l_{\tau_i}}$ follows a normalized exponential power delay profile $q^{l_{\tau_i}}=\frac{\exp(-0.1 l_{\tau_i})}{\sum_i \exp(-0.1 l_{\tau_i})}$.
The system SNR is defined as $\mathrm{SNR} = \frac{1}{N_0}$ and the pilot power is $\left|x_p\right|^2 = [10,30]$ dBw\cite{RavitejaOTFSCE}.
The DC window is designed with $\mathrm{SL}_w [\mathrm{dB}] = -40$ dB such that $k_{\mathrm{main}} \approx 3$.
All simulation results are averaged over more than $10^4$ OTFS frames.

\begin{figure}[t]
	\centering
	\includegraphics[width=3.5in]{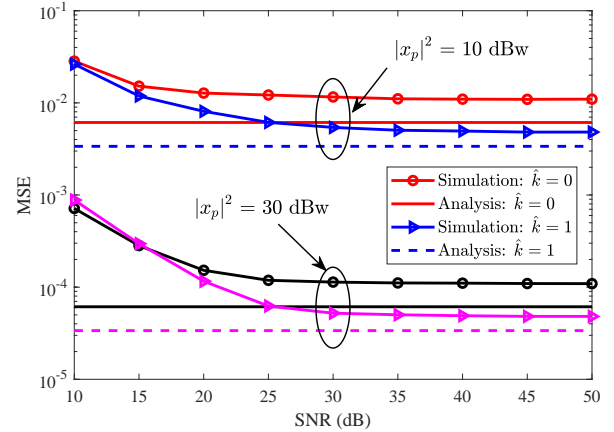}\vspace{-3mm}
	\caption{The MSE of effective channel estimation with employing a rectangular window at the transceiver.}\vspace{-5mm}
	\label{RectWindowInterspread}%
\end{figure}

Fig. \ref{RectWindowInterspread} shows the MSE of effective channel estimation performance when employing  rectangular window at the transceiver.
We can observe that the effective channel estimation suffers from an error floor in all the considered cases.
This is due to the interference spread from data symbols to the guard space caused by the existence of the fractional Doppler.
Moreover, our derived error floor level in \eqref{ChannelESTIMATIONMSE_iI} matches closely with the simulation results in the high SNR regime.
Note that a better effective channel estimation performance can be achieved with a higher pilot power.
Besides, as expected, the more additional guard space $\hat k$ inserted, the lower the MSE of channel estimation will be, at the expense of a higher amount of overhead.

\begin{figure}[t]
	\centering
	{\includegraphics[width=3.5in]{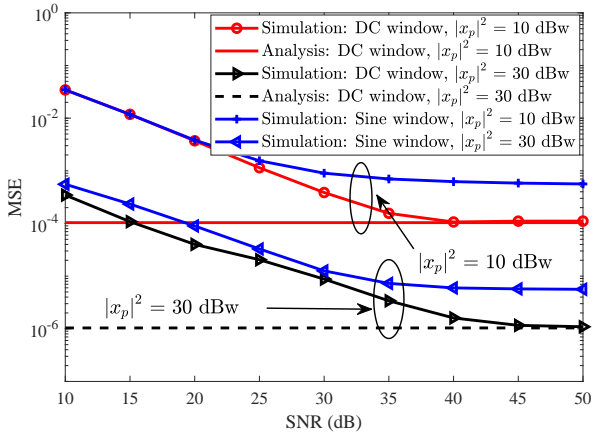}}\vspace{-3mm}
	\caption{The MSE of effective channel estimation with adopting a DC window at the transmitter and a rectangular window at the receiver.}\vspace{-5mm}
	\label{DCWindowChannelMSE}%
\end{figure}

Fig. \ref{DCWindowChannelMSE} illustrates the MSE of effective channel estimation when employing the designed DC window or Sine window at the transmitter and rectangular window at the receiver.
Note that the sine window is generated via $U_{\nu}\left[ {n} \right] = \sin\left(\frac{\pi n}{N-1}\right)$.
We can observe that the derived error floor in \eqref{ChannelESTIMATIONMSE_iI} is also consistent with the simulation results in the high SNR regime.
Comparing Fig. \ref{RectWindowInterspread} and Fig. \ref{DCWindowChannelMSE}, it can be seen that the employing the designed DC window at the transmitter is able to achieve a significantly lower MSE in the effective channel estimation than that of the rectangular and Sine windows.
This demonstrates the effectiveness of the employed DC window in enhancing the effective channel sparsity and improving the effective channel estimation performance.
We also evaluate the MSE with employing the designed DC window at the receiver and a rectangular window at the transmitter.
The results are identical to those in Fig. \ref{DCWindowChannelMSE}.
This verifies that employing DC window at either the transmitter or the receiver can achieve the same error floor for the effective channel estimation in the high SNR regime, as predicted by our analysis.

\section{Conclusions}
In this paper, we investigated the impacts of transmitter and receiver windows and proposed a window design for OTFS channel estimation.
We analyzed and revealed the insights of the impacts of windowing on the effective channel, channel sparsity, and the corresponding estimation performance.
In particular, we showed that the existence of fractional Doppler leads to the potential effective channel spread, which causes an error floor in the effective channel estimation. 
We found that adopting a window at the transmitter or the receiver can obtain an identical performance in the effective channel estimation.
Besides, we proposed to apply a DC window to enhance the channel sparsity, which improves the performance in channel estimation.
We verified the accuracy of the obtained analytical results and insights and demonstrated the substantial performance gain of the proposed window designs.

\bibliographystyle{IEEEtran}
\bibliography{OTFS}

\end{document}